\documentclass[letterpaper]{article} 
\usepackage{aaai23}  
\usepackage{times}  
\usepackage{helvet}  
\usepackage{courier}  
\usepackage[hyphens]{url}  
\usepackage{graphicx} 
\usepackage{natbib}  
\usepackage{caption} 
\usepackage{algorithm}
\usepackage{algorithmic}

\usepackage{newfloat}
\usepackage{listings}
\usepackage{booktabs}
\usepackage{amsmath}
\usepackage{amsfonts}

\DeclareCaptionStyle{ruled}{labelfont=normalfont,labelsep=colon,strut=off} 
\floatstyle{ruled}
\newfloat{listing}{tb}{lst}{}
\floatname{listing}{Listing}
\nocopyright

\title{Mastering Pair Trading with Risk-Aware Recurrent Reinforcement Learning}
\author{
Weiguang Han, \textsuperscript{\rm 1}
Jimin Huang, \textsuperscript{\rm 2}
Qianqian Xie, \textsuperscript{\rm 1}
Boyi Zhang, \textsuperscript{\rm 1}
Yanzhao Lai, \textsuperscript{\rm 3}
Min Peng\textsuperscript{*} \textsuperscript{\rm 1}
}

\affiliations{
\textsuperscript{\rm 1} School of Computer Science, Wuhan University, China\\
\textsuperscript{\rm 2} ChanceFocus AMC, China\\
\textsuperscript{\rm 3} Southwest Jiaotong 
University, China\\
\textsuperscript{*} Corresponding Author\\
han.wei.guang@whu.edu.cn
jimin@chancefocus.com
xieq@whu.edu.cn
zhangby@whu.edu.cn
laiyanzhao@swjtu.edu.cn
pengm@whu.edu.cn
}

\begin{document}

\maketitle

\begin{abstract}
Although pair trading is the simplest hedging strategy for an investor to eliminate market risk, it is still a great challenge for reinforcement learning (RL) methods to perform pair trading as human expertise. 
It requires RL methods to make thousands of correct actions that nevertheless have no obvious relations to the overall trading profit, and to reason over infinite states of the time-varying market most of which have never appeared in history.
However, existing RL methods ignore the temporal connections between asset price movements and the risk of the performed tradings.
These lead to frequent tradings with high transaction costs and potential losses, which barely reach the human expertise level of trading. 
Therefore, we introduce CREDIT, a risk-aware agent capable of learning to exploit long-term trading opportunities in pair trading similar to a human expert.
CREDIT is the first to apply bidirectional GRU along with the temporal attention mechanism to fully consider the temporal correlations embedded in the states, which allows CREDIT to capture long-term patterns of the price movements of two assets to earn higher profit.
We also design the risk-aware reward inspired by the economic theory, that models both the profit and risk of the tradings during the trading period.
It helps our agent to master pair trading with a robust trading preference that avoids risky trading with possible high returns and losses.
Experiments show that it outperforms existing reinforcement learning methods in pair trading and achieves a significant profit over five years of U.S. stock data.
\end{abstract}

\section{Introduction}
\begin{figure}[t]
		\centering
		\includegraphics[width=0.42\textwidth]{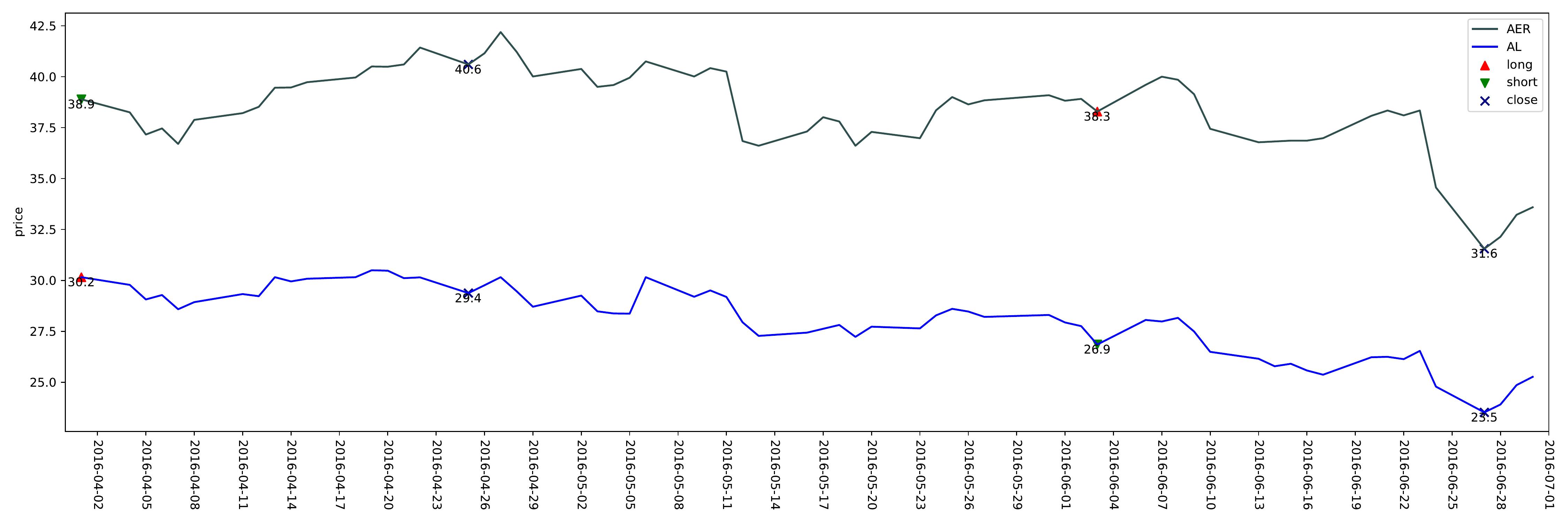}
		\caption{Pairs trading example for a pair of correlated stocks as AER and AL which shows similar co-movements. The strategy would sell the winner AER and buy the loser AL when the spread between two assets abnormally widens at first.}
		\label{pair-trading-example}
	\end{figure}

Pair trading is the simplest hedging method when an investor seeks to eliminate the market risk and has been widely adopted in the application by hedge funds.
The task consists of two steps: 1) find two correlated assets such as two stocks; and 2) trade them according to the spread refers to the difference between their prices in a subsequent period.
Therefore, it requires the strategy to precisely capture the trading opportunities when the spread of two assets abnormally widens, and earn a profit when the spread returns to its historical mean \cite{Suzuki2018OptimalPS}, as shown in Fig.\ref{pair-trading-example}.

It is still a challenge for reinforcement learning (RL) methods to perform pair trading as human expertise, although RL methods have been proven to be effective in many other areas.
The interface to employ RL in pair trading is straightforward: given two assets $\{X, Y\}$, \textbf{states} are the price features of two assets and trading information of the agent such as historical actions, cash, and present asset value; 
\textbf{actions} including \textit{long} (buy $X$ to resale it later and sell $Y$ to buy it later), \textit{clear} (no assets but cash), and \textit{short} (buy $Y$ to resale it later and sell $X$ to buy it later), can determine the status of the assets during the trading; and \textbf{rewards} are the overall trading performance of the agent on a monitoring market environment.
It is important for the agent to perform precise actions at correct trading points when there are abnormal movements of two assets.
From this perspective, pair trading provides two major challenges for RL.
\textbf{Firstly}, the overall profit relies on a series of decisions to perform correct action over hundreds and thousands of trading points, i.e, days, during the trading period.
The action on each trading point nevertheless has no direct connections to the overall profit and could have different effects considering its contextual actions.
For example, the effect of a \textit{clear} action would be different depending on whether it resides between two tradings or at the end of a \textit{long}/\textit{short} trading.
\textbf{Secondly}, acting in pair trading means to reason over infinite states due to the time-varying market, most of which have never appeared in history.
It would be difficult to detect and predict abnormal movements of two assets from all possible states, which can attribute to a number of reasons such as the breaking news of one asset, even though the fluctuations of the market are mitigated by hedging.

For these reasons, although there were previous efforts adopting RL methods to automatically perform pair trading~\cite{fallahpour2016pairs,kim2019optimizing,Brim2020DeepRL,Xu2020DynamicPM,wang2021improving,Kuo2021SolvingUL,Kim2022HybridDR,Lu2022StructuralBP,Han2023SelectAT}, it is observed that they have the tendency to perform frequent tradings with high costs and losses, resulting in a limited performance at the level of a human amateur investor.
\textbf{On the one hand}, they generally ignore the sequential information of asset prices in history, which makes it impossible to detect long-distance correlations between two actions such as opening and closing on two trading points between a series of holding actions.
Therefore, their agent can only consider the short-term pattern and perform frequent tradings for ignoring the long-term trading chances.
\textbf{On the other hand}, their agent is guided by maximizing the overall profit of the trading period.
According to the economic theory: Expected Utility theory~\cite{Fishburn1970UtilityTF} which models the decision process under uncertainty, maximizing the cumulative profit can be deemed as directly maximizing the expected utility.
However, it is reported in previous economic research that directly maximizing the expected utility can lead to irrational investment preference~\cite{Briggs2014NormativeTO}.
Their agents have the tendency to perform risky tradings with potentially high returns and losses rather than consistently profitable opportunities.

In this paper, we propose CREDIT, a reCurrent Reinforcement lEarning methoD for paIrs Trading which learns to trade like a human expert.
The first distinguishing feature of CREDIT is our employment of the bidirectional GRU~\cite{cho2014properties} and a temporal attention mechanism \cite{Desimone1995NeuralMO}, which means our method can focus on similar temporal patterns with multiple time points in history when generating actions at each point of the trading period.
It allows our method to fully exploit long-distance correlations between actions in two trading points which can indicate an abnormal spread widen.

Although this is sufficient for our agent to trade infrequently, our agent would still have the preference for risky long-term tradings which although have higher returns but also have higher losses than short-term tradings, if guided by maximizing the overall profit.
This would a major issue, especially in the time-varying future market where our agent would have potentially high risks and losses than previous methods.
To address it, we further design a risk-aware reward inspired by the Expected Utility theory, to guide the agent which considers both the profit and the risk of the whole trading period.
It forces the agent to avoid actions leading to risky tradings, i.e, losing 100\% money with one trading but earning back 150\% money with another trading, which would be deemed as the optimal actions if only the maximum profit is considered.
Experimental results on the real-world dataset over five years of U.S stock data demonstrate that our method can achieve the best performance.
Compared with previous methods, the agent in our method trades less frequently but yields a higher return.
	
In summary, our contributions can be listed as:
\begin{enumerate}
    \item We develop a reCurrent Reinforcement lEarning methoD for paIrs Trading (CREDIT), which can trade similar to a human expert rather than frequently tradings as a human amateur investor.
    \item To the best of our knowledge, this is the first attempt at pairs trading that guides the RL by a risk-aware reward based on the fully exploited temporal information from historical price features of two assets via Bi-GRU along with the temporal attention mechanism.
    It allows our method to capture the long-term profitable trading opportunities rather than short-term tradings with potential great loss and high transaction costs.
    \item Empirical experimental results conducted on stock pairs from U.S. markets demonstrate the effectiveness and superior performance of our method compared with previous RL-based pairs trading methods, which is compatible with human experts.
\end{enumerate}

\section{Related Work}
	\subsection{Traditional Pair Trading Methods}
	According to how they measure the historical correlations of two assets, traditional pairs trading methods can be divided into three different categories, including the distance metric\cite{gatev2006pairs}, the cointegration test\cite{vidyamurthy2004pairs}, and the time-series metric\cite{elliott2005pairs}.
	With $n$ assets and all possible $C_n^2 = \frac{n(n-1)}{2}$ combinations of assets, they would first select the optimal pair with the lowest measurement for a subsequent trading period.
	Then they would perform tradings when the price spread diverges more than a predefined open threshold, and closed upon mean-reversion, at the end of a trading period, or a predefined stop-loss threshold~\cite{krauss2017statistical}.
	The fixed trading strategy with predefined thresholds is model-free and easy to implement, but it would be difficult for human experts to set optimal thresholds considering the time-varying market.
	
	\subsection{Reinforcement Learning in Pairs Trading}
	Inspired by the success of applying reinforcement learning(RL) in financial trading problems~\cite{fischer2018reinforcement,almahdi2019constrained,katongo2021use,lucarelli2019deep}, there were also attempts in pairs trading which introduced RL methods to develop flexible trading agents which can learn from historical data to determine when to trade.
	The first attempt was~\cite{fallahpour2016pairs}, which used the cointegration method to select trading pairs, and adopted Q-learning~\cite{watkins1992q} to select optimal trading parameters.
	Following it, there were researches~\cite{kim2019optimizing,Kuo2021SolvingUL,Lu2022StructuralBP} further introducing stop-loss and structural break risks for better boundaries selection.
	
	Different from their methods, \citeauthor{Brim2020DeepRL} directly utilized the RL methods to train an agent for trading~\cite{Brim2020DeepRL}. \citeauthor{Xu2020DynamicPM} considered multiple asset pairs for pair trading as a portfolio and perform trading actions along with weight assignment simultaneously~\cite{Xu2020DynamicPM}. \citeauthor{wang2021improving} proposed to improve the trading performance with reward shaping~\cite{wang2021improving}. \citeauthor{Kim2022HybridDR} combined two RL networks to conduct trading actions and stop boundaries simultaneously.
	However, they failed to fully exploit temporal information in the states and consider the risk of performed tradings, leading to frequent trading with potential great losses and transaction costs.

\section{CREDIT}
\begin{figure*}[t]
		\centering
		\includegraphics[width=0.9\textwidth]{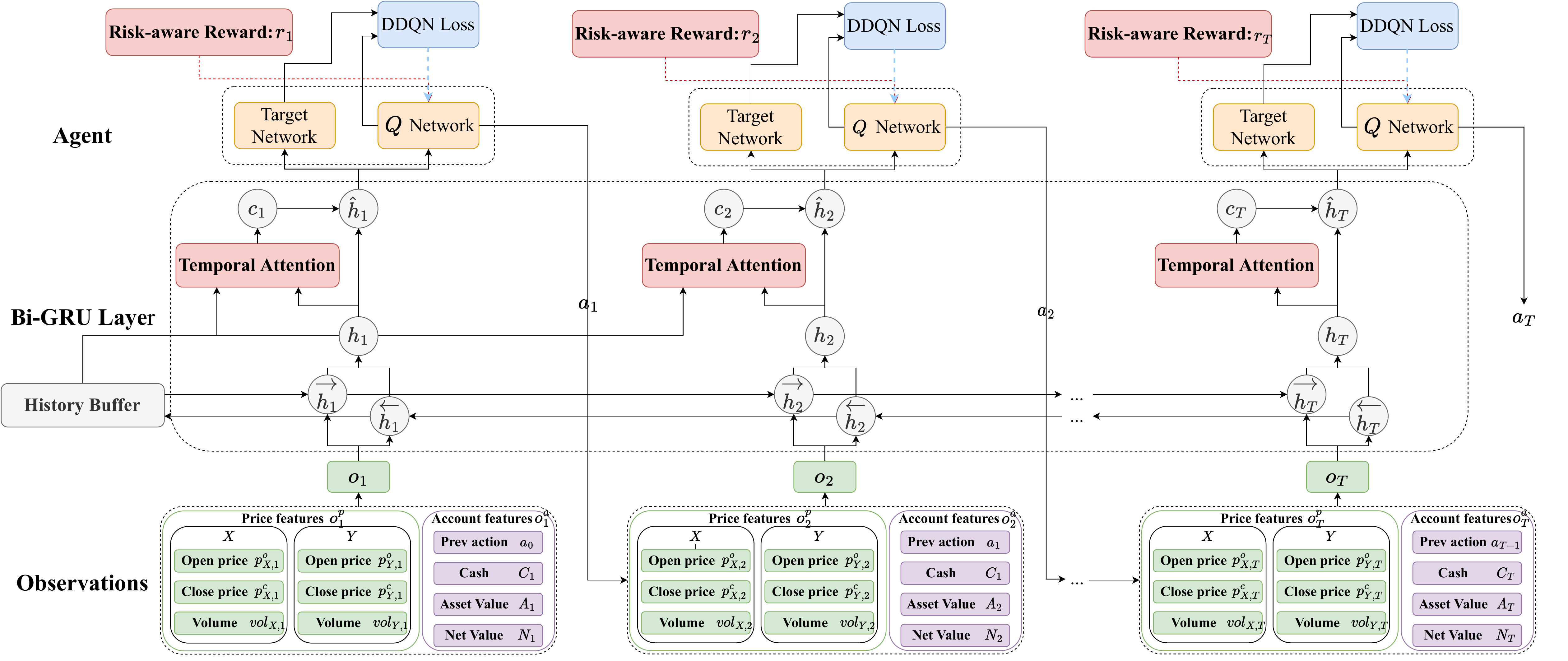}
		\caption{The structure of CREDIT.}
		\label{fig:structure}
	\end{figure*}

In this section, we present the detail of our proposed method CREDIT, as shown in Fig.\ref{fig:structure}.
\subsection{Formulation of Pair Trading}
\label{sec:reward}
\subsubsection{Preliminaries}
In this study, we mainly focus on how to trade the selected asset pair to earn market-neutral profit.
Following previous methods~\cite{fallahpour2016pairs,kim2019optimizing,wang2021improving}, we first adopt cointegration tests~\cite{engle1987co} to perform pair selection and simplify the trading from reality by requiring tradings to be performed on discrete-time, i.e, days.
Therefore, given a subsequent trading period with $T$ time points consisting of $\{0,1,\dots,T-1\}$, there are two price series $\{p_0^X, p_1^X,\dots,p_{T-1}^X\}$ and $\{p_0^Y, p_1^Y,\dots,p_{T-1}^Y\}$ of asset pair $\{X, Y\}$ that is associated with each time point.
For each asset such as $X$, the return of the time point $t$ would be:
\begin{equation}
\label{eq:asset-profit}
    r_t^X = \frac{p_{t}^X}{p_{t-1}^X} - 1
\end{equation}
\subsubsection{Partially Observable MDP}
Formally, we formulate the decision process of the trading for pair trading as a Partially Observable Markov Decision Process (POMDP)~\cite{Hausknecht2015DeepRQ} $M = (S, A, T, R, \Omega, O)$, where $S$ refers to the state space, $A$ is the action space, $T$ is the transitions among states, $R$ is the designed reward, $\Omega$ is the partial observation state which is generated from the current state $s \in S$ and action $a \in A$ according to the probability distribution $O(s, a)$.
At each time point, the agent would perform an action $a_t \in A$ under the current state $s_t$, resulting in the transition from $s_t$ to $s_{t+1}$ with the probability $T(s_{t+1}|s_t, a_t)$.
Unfortunately, in the time-varying market, the actual market states are partially observed, since the prices of assets are driven by millions of investors and macroeconomic events.
Only the historical prices and volumes of assets, along with the historical account information of the agent such as actions, cashes, and returns can be leveraged, while other information such as news, investor sentiments, and macroeconomic variables are ignored.
In detail, the agent can only receive the observation $o_{t+1} \in \Omega$ with probability as $O(s_{t+1}, a_{t})$, which requires the agent to fully exploit the historical observations up to present time point.
Notice that there exists a gap between the observation $o$ and the market state $s$, which is ignored by previous methods assuming that $o$ is reflective of $s$.

\subsubsection{Observation}
The observation $o_t \in \Omega$ consists of two different feature sets, including: (1) the account features $o^a_t \in \Omega^a$ as previous action $a_{t-1}$, present cash $C_t$, present asset value $V_t$, and cumulative profit as the net value$N_t$; (2) the price features $o^p_t \in \Omega^p$ as the open price $p^o_{i,t}$, the close price $p^c_{i,t}$, and the volume $vol_{i,t}$ of for each asset $i \in \{X, Y\}$.
We also adopt a liquidity assumption~\cite{Lesmond2003TheIN} that simplified the impact of individual tradings on the market state by appending a constant loss to each trading, since the market friction caused by individual tradings is not the main focus of our method.
Thus the action of our agent would not affect the market state and the price features of assets in our observation, which instead is embedded by the account features.

\subsubsection{Action}
Different from single asset trading and portfolio management, the actions in pair trading are associated with two assets and are limited to a pair of contradictory trading actions.
For instance, for a single asset, the trading action space can be modeled as a set with three discrete actions $A = \{1, 0, -1\}$, which refers to long (Buy the asset for sale it later), clear (Clear the asset if longed or shorted any before), and short (Sell the asset for buying it later) respectively.
By performing action, the profit of the agent could be calculated as $R_t = a_{t-1} r_t - c|a_{t} - a_{t-1}|$, where $a_{t-1} \in A$ is the previous action, $a_t \in A$ is the present action, $r_t$ is the profit of the asset, and $c$ is the transaction cost that considers both the fee and tax paid to the brokerage company and the impact to the market when a trading performed ($a_{t-1} \neq a_t$).
This is further extended to multiple assets in portfolio management with no limitations on trading,
where each action consists of a series of individual trading action on each asset $\{a_{i,t}|i \in \mathcal{S}\}$ and $\mathcal{S}$ is the selected asset sets.
The profit of the agent would be the sum as $R_t = \sum_{i \in \mathcal{S}} (a_{i, t-1} r_{i,t} - c|a_{i,t} - a_{i,t-1}|)$.
Different from single asset trading and portfolio management, to eliminate the market risk, pair trading only considers two correlated assets and requires the agent to perform contrast trading actions simultaneously for hedging.
The trading action space can be modeled as a set $A = \{long, clear, short\} = \{1, 0, -1\}$ with three discrete actions each of which involves two trading actions for two assets $\{X, Y\}$ respectively.
In detail, the \textit{long} action means to long asset $X$ and short asset $Y$ at the same time, the \textit{clear} action to clear two assets if longed or shorted any before, and the \textit{short} action to short asset $X$ and long asset $Y$.
The profit of the agent is:
\begin{equation}
\begin{split}
R_t = a_{t-1} r_{X,t} - a_{t-1} r_{Y, t} - c|a_{t} - a_{t-1}| \\
= a_{t-1} (r_{X,t} - r_{Y,t}) - c|a_{t} - a_{t-1}|
\end{split}
\end{equation}
The profit is market-neutral for hedging the return of two assets as $r_{X,t} - r_{Y,t}$,
which is required to be precisely estimated by the agent with only historical observations up to $t-1$.

\subsubsection{Reward}
Previous methods generally maximize the cumulative profit over a trading period with $T$ time points:
\begin{equation}
    R = \prod_{t \in T} (1 + R_t)
\end{equation}
However, the agent guided by maximizing the overall profit has the tendency to risky tradings with potentially high returns and losses, which is similar to human amateur investors.
There exists a gap between the agent and human expertise especially on which action they choose with partially observations and uncertain returns.

To illustrate it, we model the process based on Expect Utility (EU) theory~\cite{Fishburn1970UtilityTF}, which is an economic theory that describes how an individual as a rational person \textbf{should} decide under uncertainty.
Apart from the probability of all possible returns, it introduces the utility of each return that measures the risk preference of the return to the investor.
Giving all possible returns $\gamma \in \Gamma$, the expected utility can be represented as:
\begin{equation}
    \mathbf{E}[U(\gamma)] = \sum_{\gamma \in \Gamma} P(\gamma)U(\gamma)
\end{equation}
where $P(\gamma)$ is the probability of $\gamma$ and $U(\gamma)$ is the utility of $\gamma$.
From this perspective, maximizing the cumulative profit can be deemed as directly maximizing the expected utility.
Given all returns in the trading period ${r_t | t \in T}$ and utility function as $U(r) = ln(1+r)$, the expected utility would be $\prod_{t \in T} (1+r_t)$ which is the same as the cumulative profit.

Nevertheless, maximizing the expected utility which aims to perform rational decisions can lead to irrational investment preference, as reported in previous research~\cite{Briggs2014NormativeTO}.
For example, given two agents $A$ and $B$ both performing two tradings in the same period, $A$ first earns 100\% and loses 30\% while $B$ earns 10\% for each trading.
According to the expected utility, our method would prefer $A$ over $B$ since $\mathbf{E}[U(A)] = 2.0 \times 0.7 = 1.4$ is greater than $\mathbf{E}[U(B)] = 1.1 * 1.1 = 1.21$.
As a matter of fact, $A$ has the tendency to perform risky tradings whose potential risk is ignored in the expected utility.

The irrational investment preference can be an important issue for pair trading since the agent would be misguided to capture the fluctuations of the market rather than the price spread between two assets.
However, the market information is generally ignored in pair trading, the agent would fail to predict future market fluctuations and yield poor trading performance.

\subsection{Recurrent State Representation Learning}
Instead of vanilla Deep Q-learning Networks (DQN) in previous methods, we propose to introduce the Deep Recurrent Q-learning Networks (DRQN)~\cite{Hausknecht2015DeepRQ} into pair trading, which fully considers the sequential information of asset price features in history for market state estimation.
We first model the pair trading with the POMDP framework rather than MDP, in which an observation $o_t$ is provided for the agent consisting of the account features $o^a_t$ and the price features $o^p_t$ at each time point.
Therefore, the agent is required to estimate the latent market state $s_t$ according to the history $H_t = \{o_1, a_1, o_2, \cdots, a_{t-1}, o_t\}$.
Although the market state cannot be directly observed, the historical information embedded in $H_t$, especially the sequential dependencies can help the agent to generate better estimation.

In detail, CREDIT is a Double Deep Q-learning Networks (DDQN)~\cite{van2016deep} based method with the optimal action-value function $Q^*(o,a)$, which is the maximum expected return in the future.
In DDQN, we approximate $Q$ with a neural network $Q(o,a|\theta)$ with parameters $\theta$, and update $Q$ via optimizing the differentiable loss:
\begin{equation}
    L(\theta)=\mathbf{E}_{(o_{t}, a_{t}, r_{t}, o_{t+1}) \sim D}[(y_t - Q(o_{t}, a_{t};\theta))^{2}]
\end{equation}
where $y_t = r_{t+1} + \gamma Q(o_{t+1}, \arg\max_{a}Q(o_{t+1},a;\theta_{t});\theta^{-})$ and $\theta^{-}$ is the parameter of 
a fixed and separate target network.
However, directly taking the history as the input of $Q$ network would fail to capture the sequential connections between two trading points, leading to a limited approximation of $Q$.

Therefore, we introduce Bidirectional Gated Recurrent Units (Bi-GRU)~\cite{hochreiter1997long} along with the temporal attention mechanism to encode the history before the $Q$ network.
The previous state representation $h_{t-1}$ is deemed as the previous hidden state of the forward GRU at $t-1$, and the next state representation $h_{t+1}$ as the previous hidden state of the backward GRU at $t+1$.
Consequently, the present state representation $h_t$ can be represented as:
\begin{equation}
    \begin{split}
    \overrightarrow{h_t} = \text{GRU}(o_t, \overrightarrow{h_{t-1}}),\overleftarrow{h_t} = \text{GRU}(o_t, \overleftarrow{h_{t+1}}), h_t &= [\overrightarrow{h_t},\overleftarrow{h_t}]
    \end{split}
\end{equation}
where $h_t$ is the concatenation of the forward hidden state and backward hidden state, which allows our method to capture the temporal correlations from both directions, and $o_t = [o^a_t, o^p_t]$ is the concatenation of essential price and account information at each trading point.
For discrete variables such as previous action $a_{t-1}$ in $o^a_t$, we design an embedding layer $E_a \in \mathbb{R}^{3 \times d_a}$ respectively, where $d_a$ is the corresponding embedding size.
We also normalize the price features of assets ${X, Y}$ in $o^p_t$ by logarithm.
Our method can embed salient temporal information in $H_t$ into the state representations $h_t$ based on the observations, which provides an effective approximation of the state.
$h_t$ is further fed into $Q$-network, which helps the agent to recognize long-term opportunities between two distant trading points and perform precise actions to earn the profit.

However, Bi-GRU could suffer from the long-distance forgetting problem \cite{bahdanau2014neural}, especially for pair trading whose histories consist of thousands of trading points.
Besides, similar to human expertise, our agent should pay different attention to the previous history and focus on the most relevant sub-periods.
We further devise a temporal attention mechanism to address this issue as:
\begin{equation}
\begin{split}
		s_i^t = \frac{h_t^\top h_i}{\sqrt{d_h}}, a^t = softmax( s^t_{\cdot{}}),
		c_t = \sum_{i=1}^{t-1} a_i^t h_i
\end{split}
\end{equation}
	where $i \in [1, t - 1]$ and $a_t \in \mathbb{R}^{t-1}$.
The final state representation is $\hat{h}_t = [h_t, c_t]$ which is fed into the $Q$-network to generate the present action and value.

\subsection{Risk-aware Reward}
\begin{figure}[!htb]
	\centerline{\includegraphics[width=.35\textwidth]{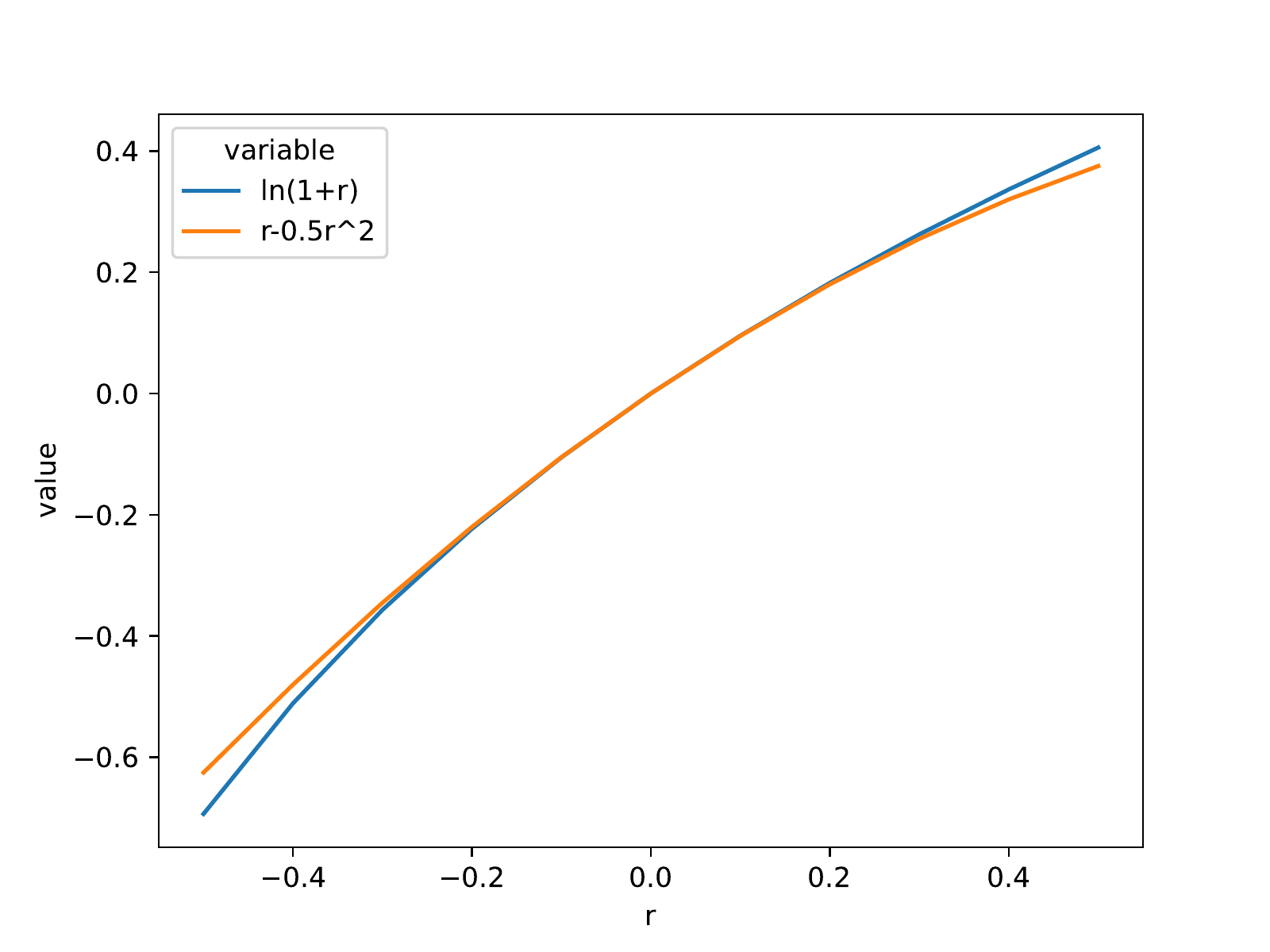}}
	\caption{The utility value of the utility function $\ln{(1+r)}$ and the quadratic approximation $r - \frac{1}{2}r^2$ under different return $r$.}
	\label{utility-example}
\end{figure}

Although CREDIT can leverage RNNs in state representation learning to fully exploit the temporal correlations embedded in history, the agent in CREDIT would pursue risky trading opportunities with potentially great loss in the future which is similar to human amateur investors, if it is guided to maximize the overall profit as previous methods.
To address this issue, inspired by previous economic research in portfolio management~\cite{Markowitz2014MeanvarianceAT}, in CREDIT, we design a risk-aware reward that incorporates the risk of trading by approximating the utility function $U(r) = \ln{(1+r)}$ with a quadratic:
\begin{equation}
    \begin{split}
    U_Q(r) = U(0) + U'(0)r + 0.5U''(0)r^2 = r - \frac{1}{2} r^2
    \end{split}
\end{equation}
As shown in Fig.\ref{utility-example}, there is insignificant difference between $\ln{(1+r)}$ and $r - \frac{1}{2} r^2$ when $-0.30 \leq r \leq 0.40$, which indicates that the quadratic approximation can be a good alternative.
Consequently, the expected value of the approximation quadratic as the reward would include the mean and variance of the returns:
\begin{equation}
R = \mathbf{E}[U_Q(r)] = E(r) - \alpha V(r)
\end{equation}
where $E$ is the mean value, $V$ is the variance, and we introduce a parameter $\alpha$.
Different from maximizing the cumulative profit, the reward considers both the profit (mean value) and the risk (variance) of the returns during the trading.
Recalling the example mentioned before in Subsection Risk-aware Reward, for two agents $A$ and $B$ ($A$ first earns 100\% and loses 30\% while $B$ earns 10\% for each trading), the expected utility of $A$ would be $0.3 - 0.49\alpha$ and $B$ be $0.1$.
Therefore, when $\alpha \geq 0.41$, our method would prefer $B$ over $A$ which is opposite from the preference of reward without considering risk, which means our reward concerns the risk of tradings more.  
This leads the agent to perform less risky tradings and more consistent performance in the future.
	
\section{Experiments}
\subsection{Data}
We select daily real-world stock data which is obtained from Tiingo End-Of-Day API\footnote{Tiingo. Tiingo stock market tools. https://api.tiingo.com/documentation/iex}.
We consider all stocks from the U.S. stock market without any missing data during the sample period, which starts on January 2, 2015, and ends on December 31, 2018.
This limits our sample to 7284 firms.
The prices of stocks are adjusted considering any corporate actions within the sample period, which is further normalized by logarithm.
Finally, as in previous methods~\cite{Brim2020DeepRL,wang2021improving,Kim2022HybridDR}, we perform pair selection using the augmented Engle-Granger two-step cointegration test~\cite{engle1987co} and select AerCap (AER)-Air Lease Corporation (AL) as the trading pair which have the lowest p-value.
The final sample consists of 1006 daily observations of the price features of two stocks.
\begin{table*}
	\centering
 \small
	\begin{tabular}{@{} c @{} c c c c c c c c @{}}
		\toprule
		Model & SR$\Uparrow$ & AR(\%)$\Uparrow$ & MDD(\%)$\Downarrow$ & AV(\%)$\Downarrow$ & AHD & TT & ABD\\ \midrule
		BAH-Long & 0.11 $\pm$ 1.07 & 6.06 $\pm$ 10.31 & 3.75 $\pm$ 0.80 & 8.22 $\pm$ 1.12 & 62 & 1 & 0 \\ \midrule
		BAH-Short & 0.08 $\pm$ 0.99 & 4.20 $\pm$ 8.91 & 3.82 $\pm$ 0.99 & 8.97 $\pm$ 1.54 & 62 & 1 & 0\\ \midrule
		CPM  & -2.02 $\pm$ 0.62 & -9.80 $\pm$ 3.59 & 3.84 $\pm$ 0.76 & \textbf{5.82} $\pm$ 0.80 & 13.22 $\pm$ 3.47 & 3.00 $\pm$ 0.50 & 9.11 $\pm$ 2.64\\ \midrule
		MLP-RL & 0.54 $\pm$ 1.12 & 7.60 $\pm$ 10.17 & \textbf{2.75} $\pm$ 1.00 & 7.46 $\pm$ 1.14 & 3.26 $\pm$ 0.31 & 14.82 $\pm$ 1.61 & 1.84 $\pm$ 0.28\\ \midrule
		CREDIT & \textbf{0.87} $\pm$ 0.74 & \textbf{9.66} $\pm$ 7.45 & 3.08 $\pm$ 0.92 & 7.90 $\pm$ 1.70 & 22.61 $\pm$ 14.24 & 10.73 $\pm$ 8.18 & 4.40 $\pm$ 5.13\\ \midrule
		w/o risk & 0.70 $\pm$ 0.65 & 9.15 $\pm$ 6.47 & 3.20 $\pm$ 0.65 & 8.47 $\pm$ 1.42 & 25.10 $\pm$ 11.60 & 4.82 $\pm$ 2.83 & 1.60 $\pm$ 1.80\\ \midrule
		w/o Bi-GRU & -0.43 $\pm$ 1.06 & 0.17 $\pm$ 7.18 & 3.38 $\pm$ 1.05 & 7.42 $\pm$ 1.83 & 29.97 $\pm$ 15.24 & 7.27 $\pm$ 6.64 & 7.52 $\pm$ 8.75\\ \bottomrule
	\end{tabular}
 \caption{Results on the AER-AL stock pair. The metric is reported as the average value of 11 rollings which is rounded to two decimals. Except for our method and baselines, we also report two ablations of our method: (1) \textbf{w/o risk} which doesn't consider the risky and only maximizes the overall profit, and (2)\textbf{w/o Bi-GRU \& temporal attention} that removes Bi-GRU with temporal attention and use the feedforward neural network to encode the history of two assets.} 
	\label{return-risk-performance}
\end{table*}

\subsection{Experiment Settings}
\begin{figure}[t]
	\centerline{\includegraphics[width=0.4\textwidth]{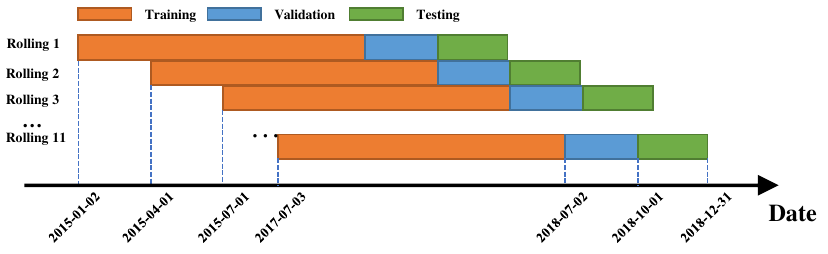}}
	\caption{The rolling and split of our experimental data.}
	\label{fig}
\end{figure}

Different from previous research, we not only split the sample period into 11 rollings, but also divide each 18-month rolling into three non-overlapping subperiods, including 12-month training, 3-month validation, and 3-month testing, as shown in Fig.~\ref{fig}.
It allows our method to select hyper-parameters for our methods and baselines according to their performance in the validation set that is unseen.
This is important since the testing data is also unseen in the application.

We compare our method CREDIT with the following baselines:
(1) \textbf{Buy and hold (BAH)}: a strong baseline that starts trading from the beginning and ends trading in the end.
(2) \textbf{Constant parameters method (CPM) \cite{fallahpour2016pairs}}: pre-defined fixed trading strategy whose trading and stop-loss threshold is set to $1.0$ and $2.0$ respectively.
(3) \textbf{MLP-RL \cite{wang2021improving}}: reinforcement-learning-based pair trading method which maximizes the overall profit with feed-forward networks.

As in previous methods, we first evaluate the performance via \textbf{return and risk} metrics, including (1) \textbf{Sharpe ratio (SR)}: the ratio of the profit to the risk~\cite{sharpe1994sharpe}, which is calculated as $(E(R_t)-R_f)/V(R_t)$, where $R_t$ is the daily return and $R_f$ is a risk-free daily return that is set to 0.000085 as previous methods. (2) \textbf{Annualized return (AR)}: the expected profit of the agent when trading for a year. (3) \textbf{Maximum drawdown (MDD)}: measuring the risk as the maximum potential loss from a peak to a trough during the trading period. (4) \textbf{Annualized Volatility (AV)}: measuring the risk as the volatility of return over the course of a year.

Different from previous methods, we are the first to employ several \textbf{trading activity} metrics to reveal the trading preference of the agent, including:
\textbf{Average holding days (AHD)}: average days of holding between opening and ending trading, or average length of trading.
(2) \textbf{Trade times (TT)}: the number of tradings during the trading period. (3) \textbf{Average empty days (ABD)}: average days between two tradings.
We report the mean value of these metrics over all rollings.
\begin{figure*}[!htb]
	\centerline{\includegraphics[width=0.9\textwidth]{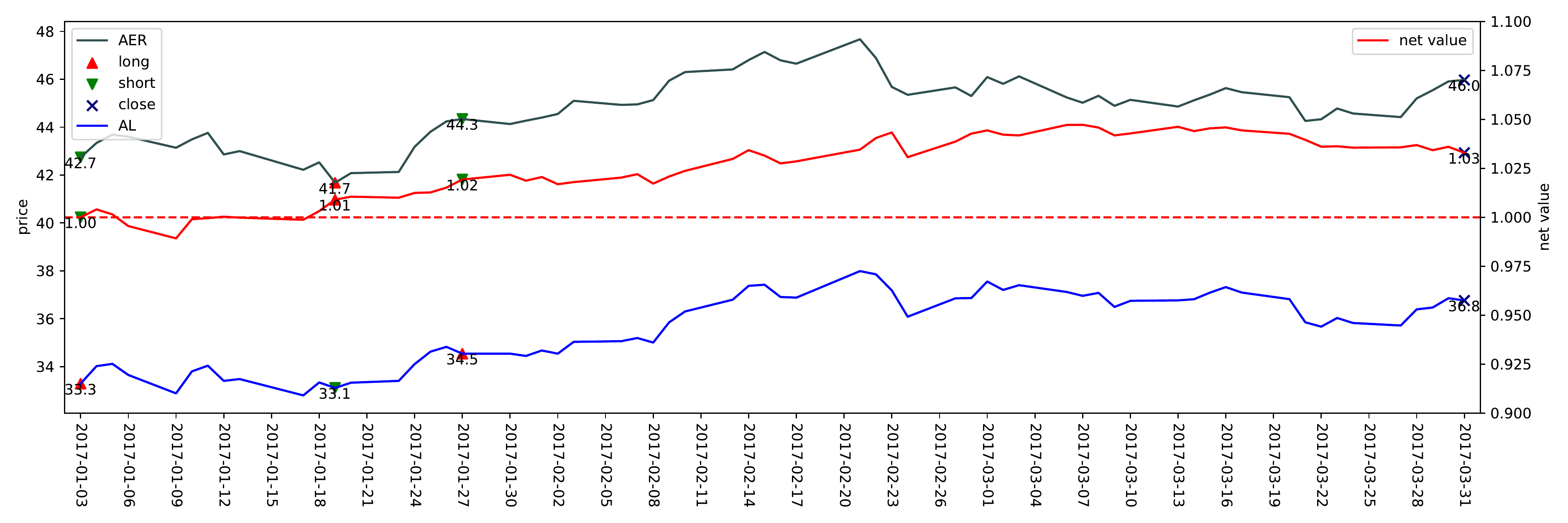}}
	\centerline{\includegraphics[width=0.9\textwidth]{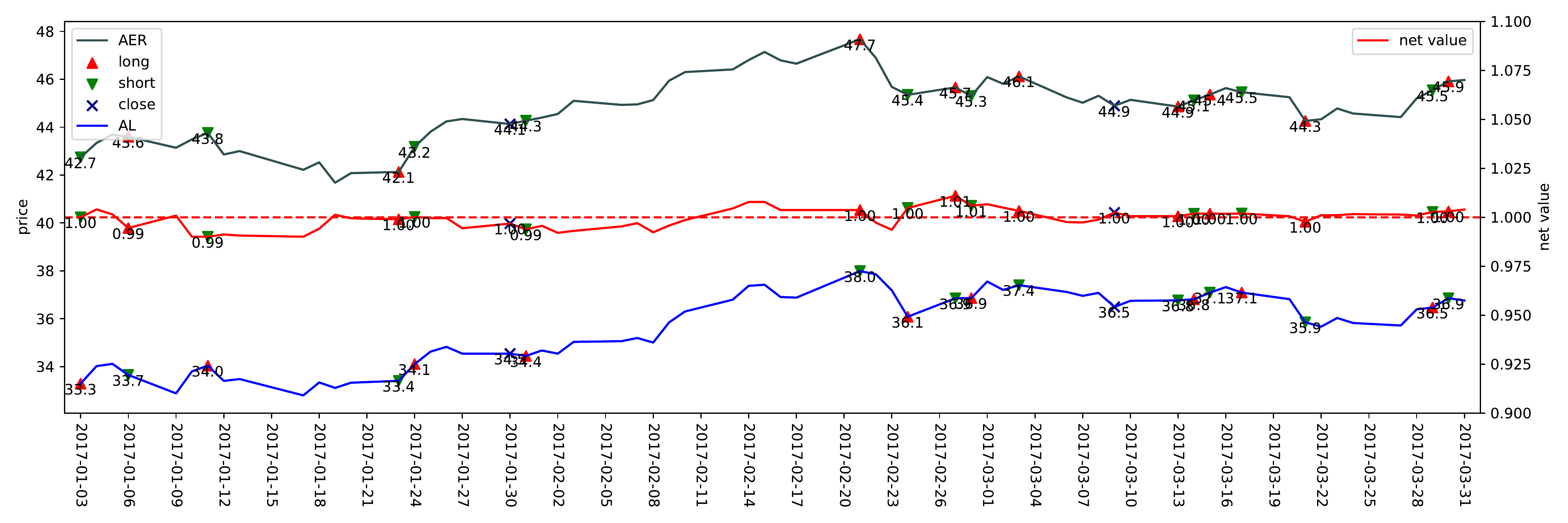}}
	\caption{Agent's trading actions on the 4th rolling of our method and MLP-RL. The above one is CREDIT and the below one is MLP-RL.}
	\label{case-study}
\end{figure*}

\section{Results Analysis}
As shown in Tab. \ref{return-risk-performance}, our method achieves the best performance over other baselines, demonstrating the effectiveness of our method.
On one hand, from the perspective of \textbf{profit} and \textbf{risk}, CREDIT achieves the highest profit with relatively low risk.
Among all methods, CREDIT can yield the highest AR and SR, which reveals that our method can effectively exploit profitable trading opportunities.
As for risk, CREDIT has the second lowest MDD which proves that our method can avoid risky tradings with large potential losses.
Our method has a relatively high annualized volatility since the metric considers the fluctuations when the return rises.
This is also proved by the significant improvement in SR of our method when compared with the RL-based method MLP-RL, since MLP-RL ignores the sequential relationships in state representation learning and the risk of each trading in reward.
Our method has a much higher AR than MLP-RL with similar MDD and AV, indicating that our method achieves remarkable profit with controlled risk.

RL-based methods such as our method and MLP-RL both outperform traditional pair trading method CPM with pre-calculated trading thresholds, which shows the essence of learning a flexible agent from history.
With wrong estimation and fixed trading rules, CPM presents the worst performance whose SR and AR are even negative.
It has inferior performance compared with BAH which performs only one trading, which means most tradings in CPM present negative returns.
In contrast, our method and MLP-RL both yields a significant profit compared with BAH.

On the other hand, from the perspective of \textbf{trading preference}, the agent in our method can perform infrequent trading with long-term holding, which is similar to a human expert.
Among all methods, CREDIT has a relatively low TT and second longest AHD and ABD, which means our agent prefers long-term trading opportunities to short-term.
In contrast, although MLP-RL can also yield a positive return, their agent has the highest TT, shortest AHD, and second shortest ABD.
This indicates that their agent performs frequent tradings during the trading most of which only holds for one day, due to their inhabit in capturing long-term correlations between historical features and the risk of the trading.
The frequent trading preference of their agent can also be observed during the trading of a human amateur investor which generally yields a limited performance.

Similar to our method, CPM has a relatively low TT and high AHD and ABD. 
However, this is due to the wrong estimation of the mean and variance of the price spread which makes it hard to trigger the trading.
Without flexible agent learning from history, the few tradings in CPM tend to cause losses rather than profit, resulting in a worse performance than BAH which only performs one trading over 62 days during the trading period.

\subsection{Ablation Study}
As illustrated by the comparisons between our method and two ablations in Tab.
\ref{return-risk-performance}, the improvement of our method compared with MLP-RL attributes to both the Bi-GRU with temporal attention and the risk-aware reward.
Our method of maximizing the overall profit (w/o risk) presents a higher MDD and AV but lower AR, which proves the essence of considering the risk of the trading in reward design.
Our method with a feed-forward neural network (w/o Bi-GRU) has a similar trading preference as our method.
However, since it cannot capture the temporal connections between two trading points, it fails to recognize profitable patterns and most tradings struggle to yield a profit, resulting in poor performance.
It clearly demonstrates the importance of integrating Bi-GRU along with the temporal attention in our method.

\subsection{Case Study}
To further demonstrate the trading preference of our agent, we also visualize the detailed tradings performed by our method and MLP-RL in the testing period of the 4-th rolling.
As shown in Fig \ref{case-study}, the price spread continues to rise in the latter 40 days of the period.
While the agent of CREDIT can effectively capture the long-term trading opportunity, the agent of MLP-RL performs a series of short-term trading which only significantly increases losses and trading costs.

\section{Conclusion}
In this paper, we present CREDIT, a reCurrent Reinforcement lEarning methoD for paIrs Trading, using recurrent reinforcement learning to dynamically learn a risk-aware agent that can exploit long-term trading opportunities as human experts.
It integrates Bi-GRU along with temporal attention mechanism to explicitly model the sequential information for state representation learning, and a risk-aware reward to guide the agent to avoid risky tradings.
Experimental results demonstrated that our method can perform long-term tradings and achieve a promising profit in the real-world stock dataset.
In the future, we will further pre-train the time series model with trading tasks and introduce trajectory-based risk indicators into the learning of the agent.
\bibliography{credit}

\end{document}